\documentclass[useAMS,usenatbib]{mnras}
\usepackage{amssymb}
\usepackage{graphicx}
\usepackage{xspace}
\usepackage{amssymb,latexsym,graphicx,natbib,eufrak,times,amsmath}

\usepackage{ifthen}
\def\draftversion{0} 

\makeatletter
\newcommand\mytoc{%
    \@starttoc{toc}%
}
\makeatother

\ifthenelse{\equal{\draftversion}{0}}{
	\usepackage{xcolor}
	\newcommand{\tmp}{}
	\newenvironment{envcomm}[1]{\renewcommand{\tmp}{#1}\begin{color}[rgb]{0,0.5,0.0}\begin{center}\hrule\vspace{0.5mm}\tmp's COMMENTS\end{center}}{\begin{center}END OF \tmp's COMMENTS\vspace{0.5mm}\hrule\end{center}\end{color}}
	\newenvironment{draft}{\begin{color}[rgb]{0,0.4,0}\begin{center}\hrule\vspace{0.5mm}DRAFT\end{center}}{\begin{center}END OF DRAFT\vspace{0.5mm}\hrule\end{center}\end{color}}
	\newcommand{\comcomm}[2]{\begin{color}[rgb]{0,0.5,0.0}\ $\bullet$ \textbf{#1:} #2 $\bullet$\ \end{color}}
	\newcommand{\revend}[1]{\par\begin{color}[rgb]{0,0.4,0}\begin{center}\hrule\vspace{0.5mm}END OF #1's REVISIONS\vspace{0.5mm}\hrule\end{center}\end{color}\par}
	\newcommand{\todo}[1]{\begin{color}{red}\ $\bullet$ \textbf{To do: }#1 $\bullet$\ \end{color}}

	\newcommand{\del}[1]{\begin{color}[rgb]{0,0.5,0.0}\ $\bullet$ \textbf{Deleted: }#1 $\bullet$\ \end{color}}
	\newcommand{\sk}[1]{\begin{color}[rgb]{0.6,0,0.6}#1\end{color}}
	\newcommand{\toc}{\par\begin{color}[rgb]{0.6,0,0.6}\begin{center}\hrule\vspace{0.5mm}\begingroup\small\let\cleardoublepage\relax\let\clearpage\relax\mytoc\endgroup\vspace{0.5mm}\hrule\end{center}\end{color}\par}
	}{
	\newsavebox{\trashcan}
	\newenvironment{envcomm}[1]{\begin{lrbox}{\trashcan}\begin{minipage}{\columnwidth}}{\end{minipage}\end{lrbox}}
	
	\newcommand{\comcomm}[2]{}
	\newcommand{\revend}[1]{}
	\newcommand{\todo}[1]{}

	\newcommand{\del}[1]{}
	\newcommand{\sk}[1]{}
	\newcommand{\toc}{}
	}

\long\def\symbolfootnote[#1]#2{\begingroup%
\def\thefootnote{\fnsymbol{footnote}}\footnote[#1]{#2}\endgroup} 

\newcommand{\eqn}[2][]{Equation#1~\ref{eqn:#2}} 
\newcommand{\fig}[2][]{Figure#1~\ref{fig:#2}}

\newcommand{\sect}[2][]{Section#1~\ref{sec:#2}}

\renewcommand{\eqn}[2][]{equation#1~(\ref{eqn:#2})}
\renewcommand{\fig}[2][]{Fig#1.~\ref{fig:#2}}

\newcommand{\bb}[1]{\ifmmode \mbox{\boldmath $ #1$} \else  \mbox{\boldmath $#1$} \fi}

\newcommand{\U}[1]{\ensuremath{\mathrm{~#1}}}

\newcommand{\pc}{\U{pc}}

\newcommand{\msun}{\U{M}_{\odot}}
\newcommand{\Msun}{\msun}

\newcommand{\ramses}{{\small RAMSES}\xspace}

\newcommand{\dx}[1]{\mathrm{d}#1}

\title[Supernovae feedback in turbulent medium]{Supernovae feedback propagation: the role of turbulence}

\author[Ohlin, Renaud \& Agertz] {Loke Ohlin\thanks{loke.lonnblad@gmail.com}, Florent Renaud and Oscar  Agertz\\
Department of Astronomy and Theoretical Physics, Lund Observatory, Box 43, SE-221 00 Lund, Sweden}

\date{Accepted 2019 March 7. Received 2019 March 4; in original form 2019 January 31}
\begin{document}
\maketitle


\begin{abstract}
Modelling the propagation of supernova (SN) bubbles, in terms of energy, momentum and spatial extent, is critical for simulations of galaxy evolution which do not capture these scales. To date, small scale models of SN feedback predict that the evolution of above-mentioned quantities can be solely parameterised by average quantities of the surrounding gas, such as density. However, most of these studies neglect the turbulent motions of this medium. In this paper, we study the propagation and evolution of SNe in turbulent environments. We confirm that the time evolution of injected energy and momentum can be characterised by the average density. However, the details of the density structure of the interstellar medium play a crucial role in the spatial extent of the bubble, even at a given average density. We demonstrate that spherically symmetric models of SN bubbles do not model well their spatial extent, and therefore cannot not be used to design sub-grid models of SNe feedback at galactic and cosmological scales.
\end{abstract}
\section{Introduction}
The conversion of gas into stars in galaxies is a highly inefficient process \citep{Kennicut98,Behroozi13, Moster13}. This inefficiency calls for some regulating mechanisms to slow down or inhibit star formation. Models of galaxy formation invoke stellar feedback to better match the observations of the stellar and gaseous contents of galaxies \citep[e.g.][]{Dekel86, White91, Hopkins14,Agertz16}, in particular in the form of SNe. SN explosions can heat and expel gas from molecular clouds, drive turbulence in the interstellar medium (ISM), and produce galactic scale outflows \citep{McKee77,Chevalier85,Strickland00,Mckee07}. It is therefore important to have a detailed understanding of how SN remnants (SNRs) propagate and affect their surroundings. 

A large number of studies have been conducted in various environments in order to understand the impact of SN explosions, ranging from ISM patches and stratified media \citep{Chevalier74,Kim01,deAvillez04,Joung06,Walch2015,Gatto17} up to galactic scales \citep{Oppenheimer06,Hopkins12,Agertz2015}. However, the important stages of the evolution of SNR occur on scaleslengths of parsecs \citep{Sedov46,Thornton98}, below what typical galactic and cosmological simulation resolve. Not properly resolving these scalelengths in numerical simulations can cause excessive cooling of the SNR, underestimating the injected energy and momentum (e.g. \citealp {Katz92}; for review, see \citealp{Naab17}). As it is computationally expensive, or even unfeasible for large galactic and cosmological simulations to reach this level of spatial resolution, sub-resolution recipes are often employed to alleviate this ``overcooling problem" \citep{Stinson06,Agertz13,Keller2014,Simpson15}. However, these models need to be tested in a controlled, small scale environment to asses how well they describe SNe at their relevant scales.

The evolution of SNRs on small scales has been studied in homogeneous media for some time, and the phases of the evolution, and their dependence on the density in this medium, are now well understood \citep[]{Ostriker88,Blondin98,Thornton98}. However, the structure of the ISM is highly inhomogeneous, and governed by turbulence \cite[see][]{Elmegreen04,Scalo04,Hennebelle12}. Therefore, how a SNR interacts with the turbulent structure needs to be understood in order to estimate not only the injected momentum and energy, but also to which scales the SNR couples to the surrounding gas. To remedy this lack of understanding, modern studies have investigated single SN explosions in inhomogeneous media \citep{Ostr15,Mart15,Walch15,Iffrig15}. These generally found that the injected energy and momentum are largely independent of the turbulent structure. However, the SNR expands faster into low density regions, making them larger than in homogeneous media \citep[but see][]{Zhang19}. The volume, energy and momentum of a SNR from these studies in inhomogeneous media have been implemented into spherically symmetric models, and used in simulations at larger scales.

However, some of these studies of single SNRs did not employ driven turbulence, instead adopting static inhomogeneities, following statistical properties of the turbulent ISM \citep{Mart15,Ostr15,Walch15}. While these simulations have the correct probability distribution functions (PDFs) and power spectra of the density, they lack the velocity structure of the turbulence. Turbulence in clouds gives rise to high density filaments \citep[see][ and references therein]{Elmegreen04,Scalo04}, and the velocity structure of these filaments may, due to the high momentum, affect the evolution of the SN bubbles. The studies implementing a velocity structure, generates the turbulence with the decay of an initial velocity field imposed on a homogeneous medium \citep{Iffrig15,Zhang19}, rather than repeated injections of energy \citep[e.g.][]{Lemaster09,Federrath10}. Because they are integrated quantities, the power spectrum and the PDF do not fully describe the detailed structure of the ISM. Therefore, a set of global statistical properties may arise from different structures, depending on how the inhomogenities are generated \citep[cf.][]{Mart15,Zhang19}.

There are studies investigating repeated SN explosions in both turbulent and stratified environments \citep{KimOstr2017,Gentry17,Fielding18}, but the fact remains that single SNRs need to be understood in order to asses their injected momentum and coupling scale to the ISM. To this end, we present numerical simulations of single SNe in a turbulent environment, representative of the ISM of local spiral galaxies. We focus on the propagation of SN explosions through this medium, investigating if the complex environment affects where the SNRs couple to the gas and inject their energy and momentum.

\section{Numerical methods}
\label{sec:Method}

We use the adaptive mesh refinement code \ramses \citep{Teyssier02} with a simulation domain of length $L= 100$ pc, and an \emph{initial} maximum resolution of $0.4$ pc.  Cells are refined when the relative difference in pressure between two adjacent cells exceeds 5\%. The domain is initialised with a uniform density of $n=100$ cm$^{-3}$ and temperature of $10$ K, using an adiabatic equation of state with a polytropic index of $5/3$. We include gas cooling, using atomic and metal line cooling tables \citep{Sutherland93, Rosen95}, at solar metallicity. Gravity is ignored as it would not significantly alter the dynamics of the medium during the short evolution of the SNRs that we are considering ($\sim 100$ kyr). 

The turbulence is generated using the forcing model of \cite{Padoan99}, which imposes a randomised forcing field between two wavenumbers $k_1=2\pi/\lambda_{\mathrm{max}}$ and $k_2=2\pi/\lambda_{\mathrm{min}}$  in Fourier space, normalised to a \citet{Kolmogorov41} power spectrum $E\dx{k}\propto k^{-7/6}\dx{k}$. In our simulations the range of driving is set to $ 1\leq k L/2\pi \leq 2 $, corresponding to scales of $\lambda_\mathrm{max}=100$ and $\lambda_\mathrm{min}=50$ pc respectively. A new forcing field is generated each turnover time, set to $2\pi/k_2\sigma_v$, where $\sigma_v$ is the requested velocity dispersion of the box, chosen as $\sigma_v =5$ km/s, in line with the \citet{Lar81} relation. While the force is not applied to all wave numbers, the turbulence naturally cascades down to small scales. The turbulent forcing corresponds to the equipartition of energy between the solenoidal and compressive modes. The turbulence heats the gas, which reaches an average temperature of $30$ K, resulting in a Mach number of $\mathcal{M}\simeq3$. In total, we run 13 different realisations of turbulence, all with the same parameters except for the seed of the random forcing.   

The turbulent forcing is driven for about two turnover times, for the ISM to reach a statistical steady state. At this time, the resolution in the central parsec is increased to $0.1$ pc. In this high resolution region, a single SN is initialised using thermal injection of the canonical energy $E_\mathrm{th}=10^{51}$ erg into one cell. No mass or metals are injected. Thermal injection methods for SNe can underestimate the momentum of the bubble.  \cite{Ostr15} demonstrated that SN cooling radii ($r_\mathrm{c}$, e.g. $\sim 1$ pc at $\sim 10^3$ cm$^{-3}$) need to be resolved by at least 3 resolution elements in order for the final momentum to converge. Our simulations satisfy this criterion at all times. At the time of the injection, the turbulent forcing is stopped in order for it not to affect the SNR. This means that the turbulence we are modelling is decaying. However, the SNR evolves with velocities on the order of $\gtrsim 100$ km$\,$s$^{-1}$, compared to the $\sim 5$ km$\,$s$^{-1}$ of the turbulence, so we do not expect the decay to have time to significantly change the medium. Along with the energy, a passively advecting scalar is added, used as a ``flag'' for further refinement (up to $0.1$ pc) within the SN bubble and around the shock front. 

To estimate more precisely the volume affected by the SNe, we ran the same simulations (e.g. same seed for the random forcing), switching off the turbulence at the same time, but without initialising the SNe (hereafter ``no-SN'' runs). This allows us to quantify the energy and momentum of the box, with and without the effect of the SN. Along with the SNe in turbulent media, we also carried out reference runs in homogeneous media of varying densities $n=1-100$ cm$^{-3}$, using the exact same methodology.

\section{Results}
\label{sec:results}

\subsection{Energy and momentum injection}
The momentum and energy of the SNRs are estimated by subtracting the total momentum, kinetic energy and thermal energy of the no-SN simulations along their evolution. As the time scales considered here are short $(\sim 100 \text{ kyr})$, the changes in both energy and momentum in the no-SN runs can be approximated using linear fits in time, giving typical errors of $\sim 10^{47}$ erg and $10^3 \Msun$ km$\,$s$^{-1}$ respectively, which are $\lesssim 1$ \% of the typical values expected in SNRs \citep{Cioffi88}. The fits are then subtracted from the total energy and momentum of the runs with SNe at the corresponding times. As the bubbles are not spherical, we compute the sum of the absolute momentum $|p|$ in each cell. Contrary to the radial momentum $p_r$, $|p|$ not only accounts for the non-spherical, asymmetric volume and expansion of the bubble, but also captures the cancellation of the SNR momentum with that of the surrounding environment.

\begin{figure}
\includegraphics{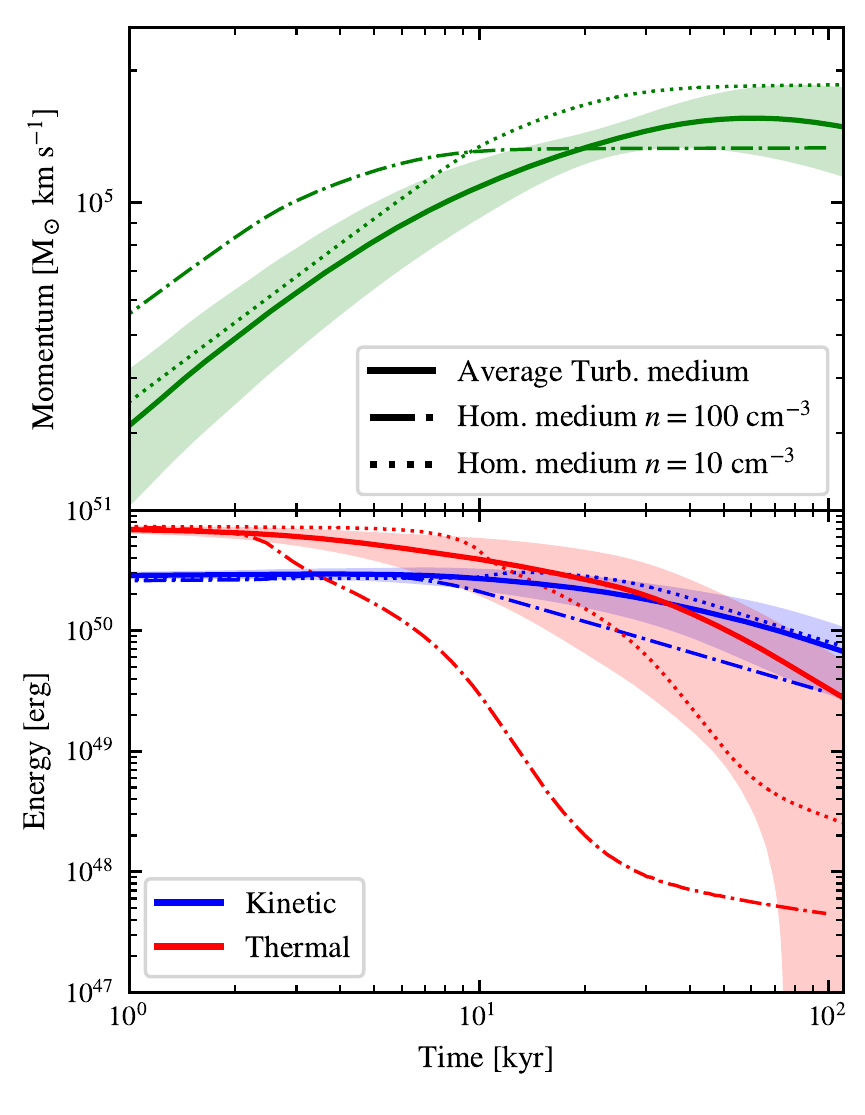}
\caption{Average evolution in time of the absolute momentum $|p|$ (upper panel), kinetic and thermal energy (lower panel) in solid lines, with the standard deviations indicated by the shaded areas. The corresponding evolution in homogeneous medium in the same average density ($n=100$ cm$^{-3}$) are shown with dash-dotted lines, and in a lower density ($n=10$ cm$^{-3}$) in dotted lines for reference.}
\label{fig:enmo}
\end{figure}

In \fig{enmo}, the time evolution of the thermal, kinetic energies and momentum $|p|$, averaged over all runs, are compared to the equivalent evolution in a homogeneous medium of $n=100$ cm$^{-3}$ (the average of the box) and $n=10$ cm$^{-3}$, for reference. The dependence of the maximum momentum on the density in the homogeneous runs is in line with expectations from analytic theory ($\propto n^{-1/7}$ \citealp{Cioffi88,Ostriker88}). The evolutions in the homogeneous and turbulent media are similar. The momentum of the turbulent gas does not appear to affect the SNR evolution until late times ($\gtrsim 100$ kyr), where momentum cancellation is visible as a decrease. At late times, the momentum is slightly higher than that found for SNRs in a homogeneous media of $n=100$ cm$^{-3}$, aligning more to the evolution in densities between $n=10$  and $100$ cm$^{-3}$. The same is found for the energies, which appear to evolve on time scales more similar to a SN in a homogeneous medium of $n \leq 10$ cm$^{-3}$, with the loss of thermal energy occurring at earlier times than $n=10$ cm$^{-3}$, but staying higher for a longer time. This implies that the SNRs couple to low density gas, allowing the bubbles to stay warmer for longer and generating more momentum than expected in a homogeneous medium.   

\subsection{Spatial evolution}
\label{spatial}

\begin{figure*}
\includegraphics{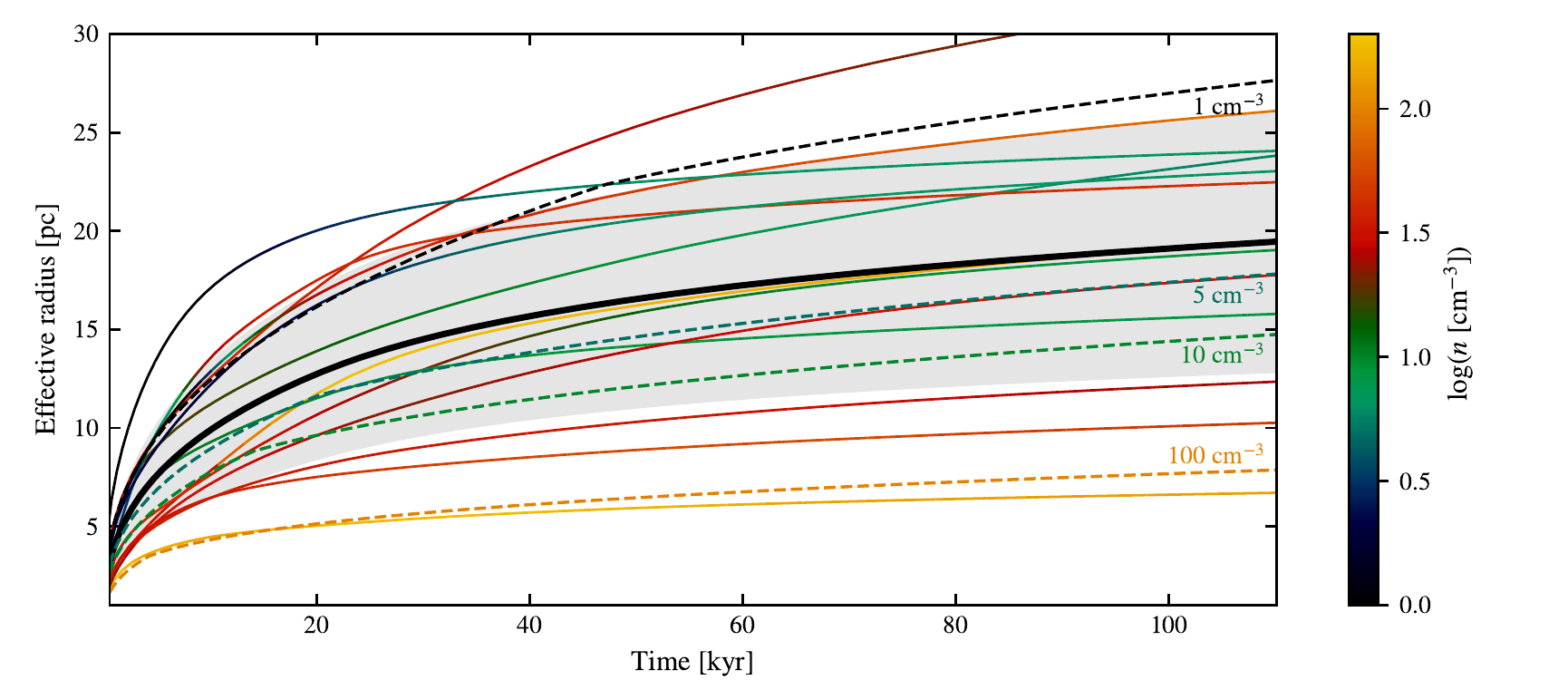}
\caption{Time evolution of the effective radius (see text) occupied by the SN bubbles. The thin coloured lines show the spatial evolution of individual SN bubbles, and the thick black line shows the average evolution for SNe in turbulent media, with the standard deviation indicated by the shaded area. The dashed lines are the semi-analytical solutions for homogeneous environments of different densities (see equation \ref{eqn:hom_rad}). The colour of the individual lines shows the average density encountered by the shock front as it expands.}
\label{fig:reff}
\end{figure*}

To estimate the size of a SNR, we consider the volume of the cells that are hotter than $10^4$ K, and define an effective radius $r_{\mathrm{eff}}$ from a sphere of equal volume. This estimate does not account for the cold shell (of which the temperature is of the same order of magnitude as the environment) that forms at late times, but only for the hot interior of the bubble. However, the shell does not represent a large fraction of the volume ($\sim 0.1\%$), as the bubble expands to a few tens parsecs, compared to the typical shell width ($\lesssim 1 \pc$). 

\subsubsection{Homogenous medium}
The sizes of SNRs are often described using the transition from the adiabatic pressure driven Sedov-Taylor stage to the momentum driven snowplough stage \citep[see][]{Cioffi88,Thornton98}, specifically through the cooling radius $r_\mathrm{c}$ and cooling time $t_\mathrm{c}$, of which exact definitions vary slightly between studies \citep[cf.][]{Mart15,Ostr15}. We define $r_\mathrm{c}$ and $t_\mathrm{c}$ as the time and radius of the bubble when the thermal energy has decreased to $3.5\times 10^{50}$ erg, i.e. roughly half of what is expected during the Sedov-Taylor stage \citep{Ostr15}. Fitting over multiple initial densities, we get relations for $r_\mathrm{c}$ and $t_\mathrm{c}$ as:
\begin{align}
    r_{\mathrm{c}}&=3.14\text{ pc}~(n/100 \,\mathrm{cm}^{-3})^{-0.42} \\
    t_{\mathrm{c}}&=3.06\text{ kyr}~(n/100 \,\mathrm{cm}^{-3})^{-0.57},   
\end{align}
in line with previous works \citep{Mart15,Ostr15}. We then fit the bubble radius as a function of time, expressed using the above expressions for $r_\mathrm{c}$ and $t_\mathrm{c}$:
\begin{align}
    r(t)=\begin{cases}
    r_\mathrm{c}\left(\frac{t}{t_\mathrm{c}}\right)^{0.38} \qquad t<t_\mathrm{c}\\
    r_\mathrm{c}\left(\frac{t}{t_\mathrm{c}}\right)^{0.25} \qquad t>t_\mathrm{c}.
    \end{cases}\label{eqn:hom_rad}
\end{align}
We use these fits for $r(t)$ as comparisons with runs in turbulent media. Both exponents in \eqn{hom_rad} are slightly smaller than the analytical values of $2/5$ ($= 0.40$) and $2/7$ ($\approx 0.29$) for $t<t_\mathrm{c}$ and $t>t_\mathrm{c}$ respectively \citep{Ostriker88,Cioffi88,Blondin98}.  However, the differences are small, and do not affect our conclusions.     

\subsubsection{Turbulent medium}
\fig{reff} shows $r_\mathrm{eff}$ from individual simulations with turbulence, as well as the average $r_\mathrm{eff}$ from all our models, alongside with the evolution in homogeneous media of various densities, taken from \eqn{hom_rad}. The average $r_\mathrm{eff}$ is similar to that in a homogeneous medium of $n=5$ cm$^{-3}$, i.e. larger than expected from the average density $n=100$ cm$^{-3}$. This can be explained by the SNRs expanding into low density escape channels (as seen in \fig{slicemap}). At late times, the slope of the average is comparable to that of the homogeneous cases. This seems to be only valid for the average, but not for individual runs in general. The reason for this could be the different realisations of high density structures. Some SNe are completely surrounded by dense filamentary, or sheet-like structures, while others are not (see examples in \fig{slicemap}).

Selecting the cells hotter than $10^4$ K, and finding their equivalents in the no-SN runs, we estimate the original average density of the volume occupied by the SNR. This is a measure of the average density encountered by the bubble, and is shown on the individual lines in \fig{reff}. While for some SNRs the expansion temporarily agrees with that in a homogeneous medium of the same density as the one encountered by the SNR, there are multiple cases where there is no such agreement. Furthermore, the cases where the bubbles slow down strongly correlate with an increase in encountered density. 

In short, the structure of the surrounding medium drastically influences the spatial evolution of SNRs. Even media with the same global properties (density, Mach number, etc.) can yield significantly different bubble sizes, due to the different realisations of the same turbulent field.

\subsection{Case analysis}
\label{sec:cases}

\begin{figure*}
\includegraphics{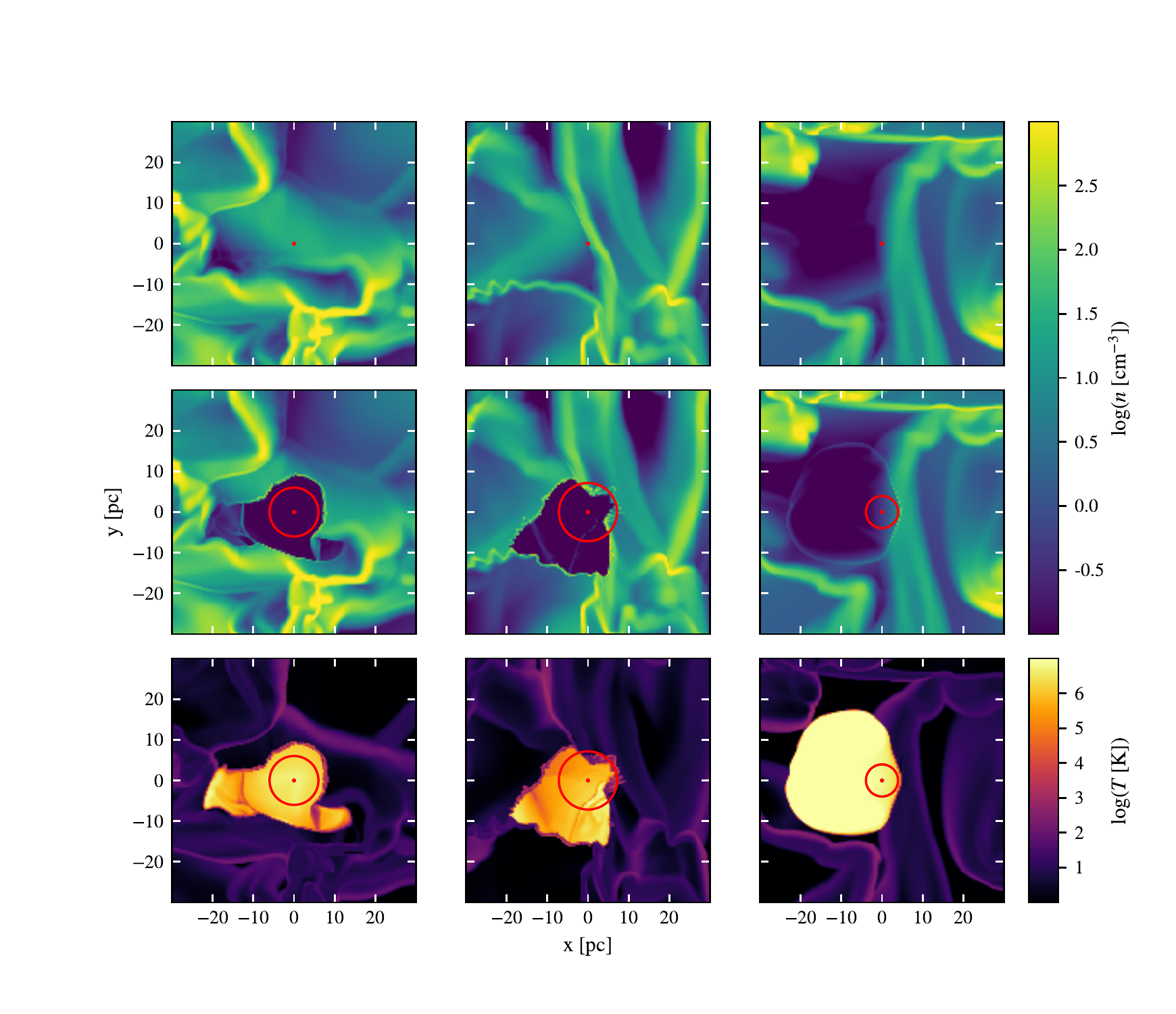}
\caption{Slice maps (with a depth of 0.4 pc) of three different runs (SN1, SN2, SN3), with the central red dot showing the position of the explosion and the red circle indicating the semi-analytical solution in an equivalent homogeneous medium (see equation \ref{eqn:hom_rad}). Top row: density of the ISM before the explosion. Middle and bottom: density and temperature when the SNRs have reached an arbitrarily-chosen volume with an effective radius $r_{\mathrm{eff}}=15$ pc (reached $37 ,75, 7$ kyr after the explosion, respectively).}
\label{fig:slicemap}
\end{figure*}

\begin{figure}
\includegraphics{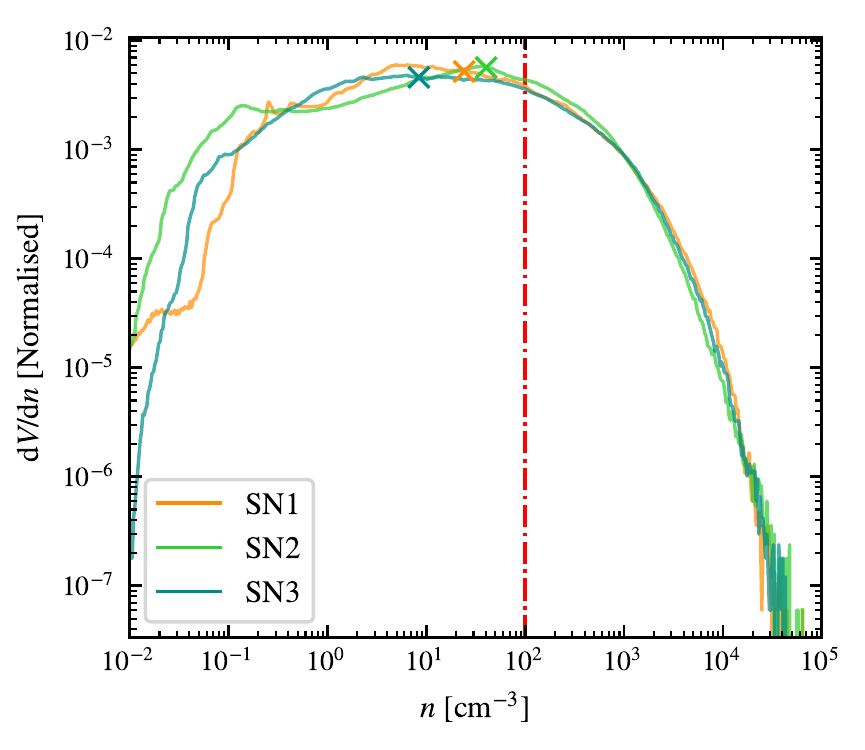}
\caption{Volume-weighted density PDFs of the SN1, SN2 and SN3 runs before the explosion, with the average density of the innermost 10 pc shown by the crosses. The dash-dotted line marks the average density of the simulation volumes (100 cm$^{-3}$).}
\label{fig:dPDF}
\end{figure}

To detail the variations between the realisations of the turbulent field, we conduct a case study on three individual SNe, labelled SN1, SN2 and SN3, chosen for illustrative purposes. \fig{slicemap} shows slice maps of the density and temperature, and \fig{dPDF} presents density PDFs before the SN explosions, with indications of the local densities of the SNe, defined as the average density withing $10$ pc of the explosion site. The cases differ in \emph{local} density by less than a order of magnitude, with values of $24,~40,~8$ cm$^{-3}$ respectively. These realisations showcase three different density structures, with SN2 and SN3 exploding next to a filament, but SN3 having a large nearby region of low density to expand into. However, despite all cases having similar average local densities, they do not reach our arbitrarily-chosen size of $r_\mathrm{eff}=15$ pc at the same time ($37,~75,~7$ kyr respectively), confirming the results of \fig{reff}. 

The evolution of the SNRs reflects their local density structures, with asymmetries shown in all cases. These SN bubbles expand faster than expected from the evolution in a homogeneous environment due to low density channels near the explosions. Wherever the SNe encounter higher densities, the expansion slows down significantly, causing the bubbles to be shaped by the filaments. 

Density inhomogenities along the shock front imply different cooling rates at different locations in the bubble shell. This effect is particularly visible in the SN3 case (right column in \fig{slicemap}) where the right-hand side of the bubble encounters a dense filament, cooling it more efficiently than the rest which remains in the Sedov-Taylor phase. Because SN1 and SN2 have evolved for longer times (30 and 68 kyr more than SN3, respectively), they formed the cold shells typical of the snowplough phase \citep{Ostriker88}. The collisions of these shells and filaments have caused reverse waves to propagate inwards, cooling the bubble interiors and thus tending to even out the pressure between the bubbles and their exteriors. The increased cooling in the interiors accelerates the transition into the momentum-driven stage.

To study the asymmetric expansion of the SNRs, we seek the spatial distributions of momentum and mass, and the azimuthal profile the bubbles.
Using the previous identification method of the bubble (i.e. solely based on temperature) would miss significant fractions of the mass and momentum (which reside in the cold shell, at the same temperatures as the ambient medium). Therefore, we identify the shell as the cold medium ($<10^4$ K), containing the passive scalar used to increase the resolution (see \sect{Method}), and of which velocity has more than doubled since the explosion. We add this volume to the warm bubble interior previously identified. The difference in mass between this method and an eye-estimate is on the order $\sim 100$ M$_\odot$, i.e. $\sim 1$ \% of the total mass of the bubble.

\begin{figure}
\includegraphics{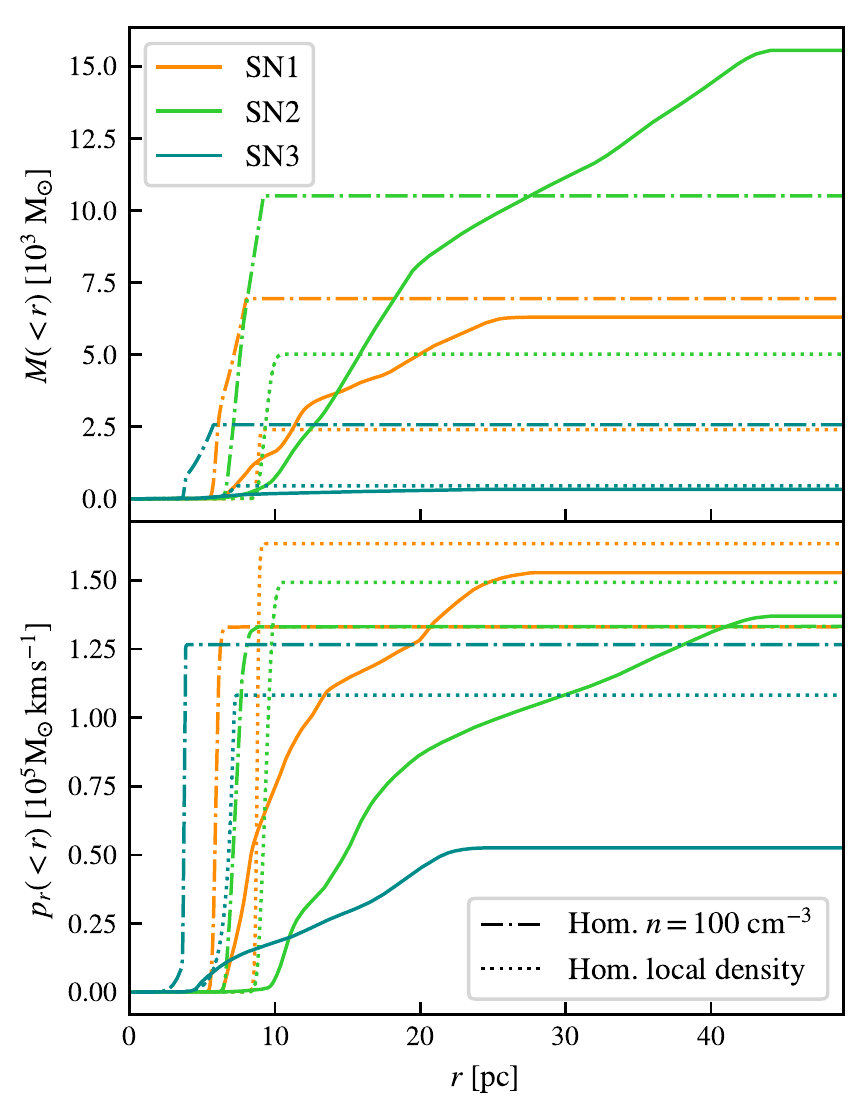}
\caption{Cumulative mass $M(<r)$ (top panel) and radial momentum $p_r(<r)$ (bottom panel) profiles of the SN bubbles, with solid lines showing each of the SNe in turbulent media. Dash-dotted lines represent the equivalent quantities for SNe in a homogeneous medium of $n=100$ cm$^{-3}$ and dotted lines the average local density, both measured at the same times as each respective SN (37, 75, 7 kyr respectively).}
\label{fig:cumsum}
\end{figure}

\fig{cumsum} shows the cumulative radial profiles of mass (top) and radial momentum (bottom). For comparison, the equivalent profiles at equivalent times for SNe in homogeneous media, both of the average local density and total average density are also shown. The mass profile of the SNe bubbles varies between the cases, not only in comparisons with each other, but also relative to the equivalent homogeneous cases. The total mass of SN3 agrees well with its homogeneous analogue at the same local density, while SN1 agrees more with a homogeneous medium of $n=100$ cm$^{-3}$, and SN2 is greater than both. This reflects the influence of the local environments as discussed above.

The cumulative momentum (\fig{cumsum}, bottom panel) also reveals differences between the various cases. SN3 has a lower momentum than both SN1 and SN2, and its homogeneous equivalences, indicative of it being in its early evolution (as already hinted by the lack of a cold shell). On the other hand, both SN1 and SN2 match the expected momenta in homogeneous environments of the same average \emph{local} densities.

\begin{figure*}
\includegraphics{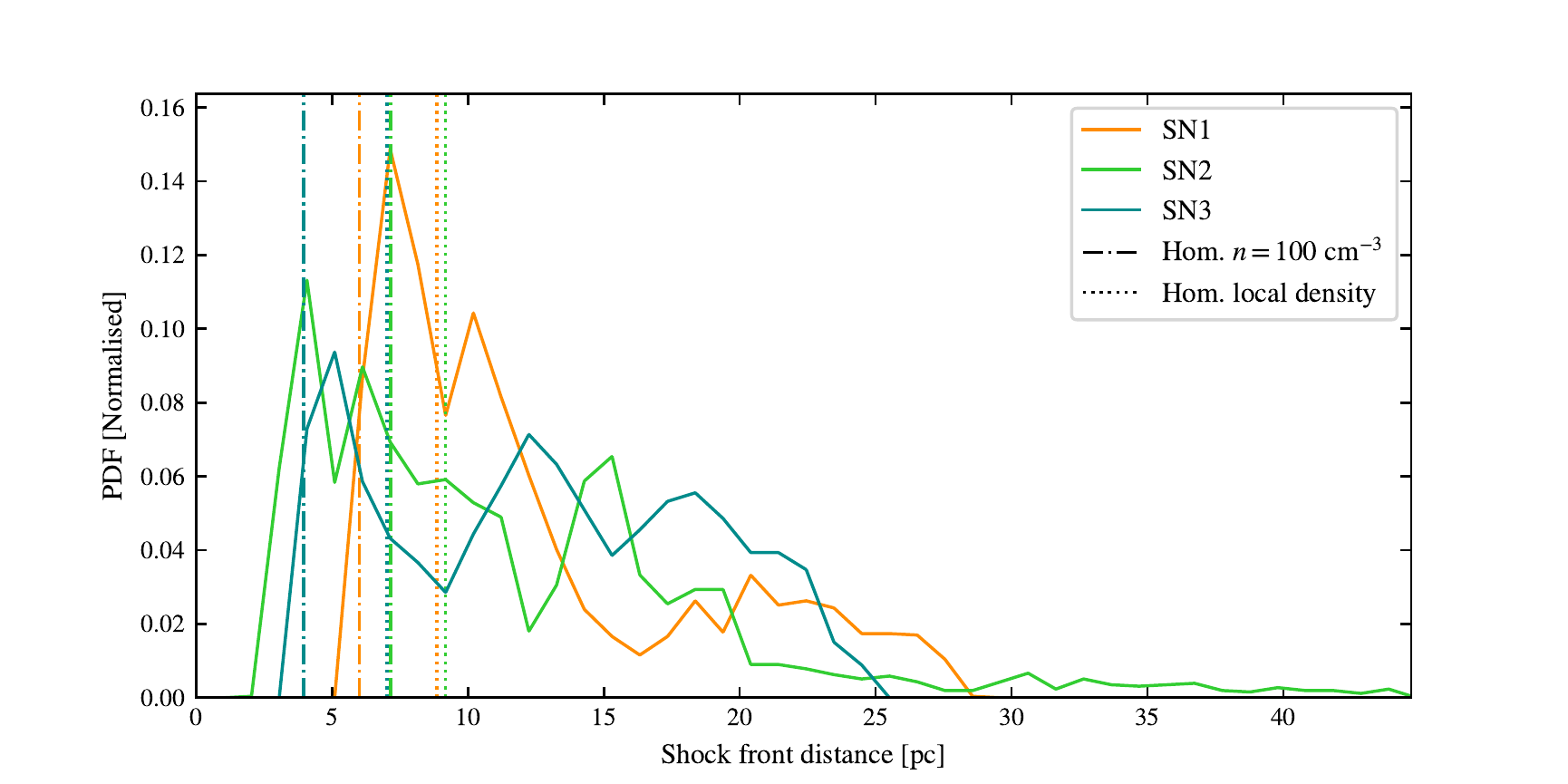}
\caption{Distribution of maximum shock front radii $r_{\mathrm{sh}}$ measured in azimuthal bins of $5^\circ \times 5^\circ$ from the explosion site. The different colours indicate different runs, at the times when the effective radius is $r_{\mathrm{eff}}=15$ pc. The vertical lines mark the semi-analytical solution for homogeneous media (see equation \ref{eqn:hom_rad}) in densities of $n=100$ cm$^{-3}$ (dash-dotted lines) and of the average local density within 10 pc of the explosion site (dotted lines).}
\label{fig:raddist}
\end{figure*}

To quantify the asymmetry of the bubbles, \fig{raddist} shows the distribution of shock front distances to the explosion site, measured in $5^\circ \times 5^\circ$ azimuthal bins. While some of the peaks in $r_\mathrm{sh}$ are close to the solution from homogeneous media, all distributions span a wide range if radii, both smaller and larger than the homogeneous solutions. In some of the cases (SN1 and SN3) the distribution at large radii is mostly clustered at specific regions (as in $\sim 10$ and $\sim 20$ pc for SN1), while for SN2 the radii extend beyond 40 pc, suggesting a escape channel not visible in the slice maps of  \fig{slicemap}.

\fig{cumsum} and \fig{raddist} show most of the mass and momentum are found at the peaks in \fig{raddist}. However, a significant fraction can be found at larger radii, especially for SN2 where $\sim 30$ \% of the mass and momentum are found in the escape channel ($\gtrsim 18$ pc). While most of the momentum is absorbed by the filaments, a sizeable fraction couples to the low density gas, and is spread over a large range of radii. This variety of radial distributions of injected momentum and swept up mass set the ability of SNe to drive galactic winds, and the mass loading these winds.

 In conclusion, the density structures around the SNRs affect not only their morphologies and sizes but also impact the mass and momentum distributions of the SNRs. None of these aspects are correctly captured by spherically symmetric models, often assumed in sub-grid recipes.

\section{Discussion}
In this study, we have only used one set of parameters, i.e. the same density, Mach number and driving scale of the turbulence for all simulations. Increasing the Mach number widens the density PDF \citep{Vazquez-Semadeni94} and thereby enhances the contrasts between the dense filaments and low density volumes. The driving scale of turbulence further affects to what extent escape channels exist \citep{Mart15}. Our study demonstrates the importance of such escape channels on setting the morphologies and extents of SNRs. Therefore, considering a wider range of parameters (average density, Mach number) would likely lead to an even greater diversity of bubbles, further impairing spherically symmetric models. 

Conversely, \cite{Zhang19} reported universal, close to spherical shapes of their SNRs, in all the turbulent media they modelled. This discrepancy with our conclusions and those of \citet{Mart15} probably originates from the differences between their turbulent structures and ours. The distribution of dense structures in their media is significantly more isotropic than in our simulations, causing the evolution of their SNR to be more spherically symmetric. This is likely due to their turbulence resulting from the decay of a one-off generated random velocity field, as opposed to our media originating from a series of forcings over several turnover times (recall \sect{Method}). We also note that the scales considered in \citet{Zhang19} are smaller than that of this study. While the initial velocity dispersions in our simulations are in line with \citet{Lar81} scalings, the higher values they use cause their SNRs to encounter more energetic structures as they expand. In addition to the isotropy aspect noted above, this could explain the differences between their conclusion on the sphericity of the bubbles and ours on asymmetric SNRs.

Pre-SN feedback such as stellar winds, radiation pressure and photo-ionisation are not accounted for in our study. These processes can significantly alter molecular clouds by creating low density cavities around the progenitor stars, and/or smoothing out density contrasts \citep{Murray09,Fall10,Dale12}. By changing the distributions of density peaks and low densities channels, these processes would influence the spatial and temporal evolution of SNRs. The bubbles would then likely expand to larger distances and, in the extreme case of a complete disruption of the cloud by pre-SN feedback, would yield more spherically symmetric shapes (due to the absence of remaining filaments).

Our simulations only consider the energy and momentum injection from a single supernova. However, stars form in clusters \citep{Lada03} and the collective feedback effects of multiple stars, with different masses and timescales, would have different imprints on the ISM. Specifically, sequential SNe highly correlated in space and time interact, forming a super-bubble \citep{McCray79,Tomisaka81,KimOstr2017,Gentry17}. Recently \cite{Fielding18} investigated the breakout of multiple SNe in a stratified disc environment, and also in a turbulent box similar to ours. Their findings indicate that multiple SNe can generate galactic outflows of the order of the star formation rate, thus regulating star formation at galactic scales. Furthermore, in \cite{Fielding18}, clustered SNe can destroy the filamentary structures of clouds, something that single SNe are inefficient at for the specific densities explored in this study. The conclusions of our work may therefore change significantly with the introduction of clustered SNe.

\section{Conclusion}
We present numerical simulations of SNRs in turbulent environments, studying whether the turbulent structures and kinematics influence the evolution of supernovae bubbles. Comparing with the evolution of SNRs in homogeneous media, our main findings are:
\begin{itemize}
    \item In agreement with previous works \citep{Ostr15,Mart15,Haid16}, the total momentum and energy injected by SNe are largely unaffected by the heterogeneous density structures. The momentum in high density filaments only has a noticable effect on the SNR at late times, when it cancels that of the bubble.
    \item Because most of the mass is concentrated in a few dense filaments, low density escape channels exist, allowing the SNRs to expand faster, and grow larger in the turbulent environment, as compared to SNe in homogeneous densities.
    \item We find that the volume of the bubbles deviates significantly from any semi-analytical solution. This scatter in bubble sizes is a direct result of the different realisations of the turbulent ISM. The volumes of the bubbles (or equivalently their filling factors) cannot be modelled without resolving the details of the filamentary structure of the ISM.
    \item While a rapid expansion into low density regions would suggest outflows, we find that most of the momentum and mass of the bubbles are found at small radii, where the bubble stalls against high density filaments. However, this can also vary between realisations, as some cases have a significant ($\sim 30\%$) fraction of their momentum in escape channels.
\end{itemize}

Small scale, cloud simulations (like this study), do not self-consistently model the full dynamical range of the ISM turbulence, where the injection scale is $\gtrsim 100$ pc \citep{Agertz09,Renaud13,Falceta-Goncalves15,Grisdale17}. As the turbulence cascades and other galactic scale mechanisms (e.g. shear, tides, shocks) affect the structure of the ISM, neglecting or over-simplifying some of these aspects might bias quantitatively the conclusions drawn here and in comparable works. This dependence on kpc-scale (hydro)dynamics varies with the location in the galaxy in a complex manner which cannot easily be modelled self-consistently in simulations of isolated clouds.

The interactions between the expanding SNRs and the dense filamentary structure of the ISM have implications on the ability of SNe to drive galactic winds. As SNRs mostly expand through low density channels, they only plough a small amount of gas. The high density medium remains within the cloud, and thus the resulting mass loading factor of these winds is low. This dependence on the detailed structure of the ISM cannot be captured by spherically symmetric sub-grid models, commonly used in galactic and cosmological simulations where these scales are not resolved \citep{Agertz13,Hopkins14,Vogelsberger14,Schaye15}. These sub-grid recipes may thus misrepresent the mass-loading factor of potential outflows by coupling energy and momentum (incorrectly) to too much gas mass, as well as the wrong gas phase. This in turn may lead to overestimating the ability of SNe to drive galactic outflows and regulate star formation. In order to fully capture the evolution of SNe, the ISM needs to be resolved on sub-parsec scales. While it might not be practical for large scale galactic and cosmological simulations, our results show that the use of sub-grid models introduce errors that are difficult to quantify and correct for.

\section*{Acknowledgements}
We thank the anonymous referee for their constructive report, and Troels Haugb\o lle and Paolo Padoan for their help with the turbulence module. OA and FR acknowledge support from the Knut and Alice Wallenberg Foundation. OA acknowledges support from the Swedish Research Council (grant 2014- 5791).




\bibliographystyle{mnras}
\bibliographystyle{mnras}

\end{document}